\documentclass[twocolumn,showpacs,aps]{revtex4}

\usepackage{amsmath}
\usepackage{amssymb}
\usepackage{color}
\usepackage{verbatim}
\usepackage{array}
\usepackage{graphicx}
\usepackage{epsfig}
\usepackage{bm}
\usepackage[ragged]{sidecap}

\newcommand{\ds}{\displaystyle}
\newcommand{\ddsum}[1]{{\displaystyle \sum_{ #1 }}}

\newcommand{\supercomas}[1]{``#1''}

\def\bra#1{\mathinner{\langle{#1}|}}
\def\ket#1{\mathinner{|{#1}\rangle}}
\newcommand{\braket}[2]{\langle #1|#2\rangle}

\newcommand{\dd}{\mathrm{d}}
\newcommand{\ti}{~}
\newcommand{\bn}[1]{\mathbf{#1}}

\begin{document}
\preprint{\hfill\parbox[b]{0.3\hsize}{ }}

\title{Angular and polarization correlations in two--photon decay of hyperfine $2s$ states in hydrogenlike ions}

\author{F. Fratini$^{1}$\footnote{
E-mail address: fratini@physi.uni-heidelberg.de}
}
\affiliation
{\it
$^1$
Department of Physics, Post Office Box 3000, FI-90014, University of Oulu, Finland\\
}

\begin{abstract}
The angular and polarization properties of the photon pair emitted in the two--photon decay of $2s$ hyperfine states in 
hydrogenlike ions are investigated within the relativistic Dirac framework and second order perturbation
theory. The studied transitions are $2s_{1/2}\,(F=1,0)\to1s_{1/2}\,(F=1,0)$ in Hydrogen atom and
$2s_{1/2}\,(F=4,3)\to1s_{1/2}\,(F=4,3)$ in hydrogenlike Uranium ion. Two different emission patterns are found:
For non-spin-flip transitions, 
the angular correlation ($i.e.$ the angular distribution of the emitted photons) is of the type $\sim 1+\cos^2\theta$, while the linear polarizations of the emitted photons are 
approximately parallel one another;
For spin-flip transitions,
the angular correlation is of the type $\sim 1-1/3\cos^2\theta$, while the linear polarizations of the emitted photons are
approximately orthogonal one another.

\end{abstract}


\maketitle

%
%
%
%

\section{Introduction}

The theoretical formalism of the two--photon decay in atoms and ions has been introduced by Max Born's 
PhD student Goeppert-Mayer in 1931 \cite{GoM31}.
Many aspects of such a process,
like the total decay rate and the spectral distribution, have been extensively 
investigated in the context of few-electron atoms and ions, both in theory and experiments 
\cite{Sh58, DrG81, Dra86, Go81, DeJ97, SaP98, DuB93, AlA97, ScM99, MoD04, KuT09, Se10}.
Recently, some interest
has been also devoted to the angular and polarization correlations of the two emitted photons \cite{An05, Les06, An09, An10, PolPol},
following the pioneer semi-relativistic work of Au \cite{Au76}.
Since the beginning of these studies, two--photon transitions in atoms and ions have been not only 
a challenging process to theoretically and experimentally investigate but also a 
useful tool with which to explore different physical areas. 
Already in 1940, for instance, Breit and Teller derived that the
double photon emission was the principal cause of the decay of interstellar hydrogen atoms from their metastable 2s state \cite{Br40}, while,
more recently, polarization correlations of the emitted photons have been employed to successfully test violations of the 
Bell inequality  \cite{AsD82,PiK85,KlD97,QuantInf}. 
Furthermore, two-photon transitions have been proposed as a tool to
measure weak interaction properties \cite{Sch89, PNC}. \\
In the present contribution, the angular and polarization correlations of the photons emitted in
two--photon decays of $2s$ hyperfine states in hydrogenlike ions are presented.
By comparing with previous theoretical results, 
where the two--photon decays of bare electron states in 
hydrogenlike ions have been analyzed,
our calculations show the impact of the nuclear angular momentum (spin)
on the angular and polarization properties of the emitted photons.\\
Nuclear features which are not easily accessible via bare nuclear physics
might be successfully explored by accurately studying nuclear effects on atomic processes.
A brilliant review 
of such effects, at the borderline between atomic and nuclear physics,
has been recently written by P\'alffy \cite{Pa210}. Along this line, owing to the
significant progress that has been recently made in the development
of position-sensitive photon detectors \cite{RSI79, JI5, RSI67, PRL97},
the present paper may also outline an alternative route to probe nuclear spin properties by using 
two--photon transitions. \\
The paper is structured as follows. In Sect.~\ref{sec:theory} the theory needed for the analysis will be explained.
Particularly, in Sect.~\ref{sec:states} the states of the overall --electron plus nucleus-- system will be defined. In 
Sect.~\ref{sec:second}, 
a detailed formulation of the two--photon bound--bound transition amplitude will be presented, within the framework 
of second order perturbation theory. Then, in Sect.~\ref{sec:ddr}
the differential decay rate and the angular correlation of the photons
will be defined in terms of the transition amplitude. In Sect.~\ref{sec:results}, results for
the angular correlation function, for the transitions 
$2s_{1/2}\,(F=1,0)\to1s_{1/2}\,(F=1,0)$ in Hydrogen atom as well as for
$2s_{1/2}\,(F=4,3)\to1s_{1/2}\,(F=4,3)$ in hydrogenlike Uranium (${}^{235}_{92}$U) ion,
will be presented and discussed. The polarization properties 
of the emitted radiation will be also analyzed in detail for a few of these transitions.
Finally, a brief summary will be given in Sect.~\ref{sec:summary}.
%
%
%
%
%
%
%
%
\section{Theory}
\label{sec:theory}

\subsection{Construction of the overall set of states}
\label{sec:states}
The presence of the nuclear spin has a twofold effect on the states of the overall hydrogenlike system. First,
the energies of the ionic metastable states are slightly shifted, mainly due to
the magnetic dipole interaction that nucleus and electron
experience. This energy correction can be described by using first 
order perturbation theory with additional 
contributions, such as the
relativistic, Bohr-Weisskopf, Breit-Rosenthal and QED contributions \cite{Sh97, Sa08}. 
Since this energy correction is too small to affect considerably the angular and
polarization properties of the emitted radiation, it will be totally neglected
in the following.\\
Secondly, the ionic states acquire a new quantum 
number, usually denoted by $F$, that represents the total angular momentum of the overall --nucleus plus electron-- system.
The overall ionic state can be thus described, to a first approximation, 
by coupling the nucleus and electron angular momenta:
\begin{equation}
\begin{array}{l}
\ket{n, \beta; F, I, \kappa, m_F}=\\[0.4cm]
\qquad \ddsum{m_I,\,m_j} \braket{j,m_j,I,m_I}{F,m_F} \ket{n; \kappa, m_j}\ket{\beta; I, m_I} ~,
\end{array}
\label{eq:totstate1}
\end{equation}
where $n$, $\kappa$ and $j$ are the (Bohr) principal, 
the Dirac and the angular momentum quantum number of the electron respectively, while
$I$ represents the nuclear spin.
Finally, $m_I$, $m_j$ and $m_F$ are the projections of the nuclear,
electronic and total (nucleus plus electron) angular momentum onto the quantization axis, respectively.
Using standard notation, $\braket{j,m_j,I,m_I}{F,m_F}$ are Clebsch-Gordan coefficients. \\
In writing Eq.\ti(\ref{eq:totstate1}), we neglected any hyperfine interaction between nucleus and electron which
could, in principle, be responsible of mixing ionic states with different quantum numbers. Such a mixing
could in turn affect the angular and polarization properties of the emitted radiation, which are
the physical quantities we are going to investigate.
Since Eq.~\eqref{eq:totstate1} will be used in what follows, we need to estimate this approximation.
We may evaluate the goodness of Eq.\ti(\ref{eq:totstate1}) by calculating
the admixture of the state $nS_{1/2}(F')$ that the state $mS_{1/2}(F)$ acquires, using the non-relativistic Hamiltonian
which accounts for nucleus-electron hyperfine interaction \cite{JB}:
\begin{equation}
\begin{array}{c}
\ds\mathcal{C}\approx k_0 \frac{8\pi}{3} \frac{\bra{n, \beta; F', I, -1, m_F}
\delta(\hat r) \hat{\bn S} \hat{\bn I} 
\ket{m, \beta; F, I, -1, m_F}}{E_{mS_{1/2}}-E_{nS_{1/2}}} \\[0.4cm]
\ds\approx k_0 \frac{8\pi}{3}\frac{\hbar^2}{2}\frac{ F(F+1) - I(I+1) -\frac{3}{4}}{E_{mS_{1/2}}-E_{nS_{1/2}}}\,
\delta_{F,F'}\,\frac{Z^3}{\pi a_{0}^3 (n\,m)^{3/2}}
\end{array}
\label{eq:ev}
\end{equation}
where $a_0$ is the Bohr radius, $k_0=2g_N\mu_0\mu_B\mu_N/4\pi\hbar^2 $ and $\mu_0$, $\mu_B$, $\mu_N$, $g_N$ are the
magnetic permeability of vacuum, the Bohr magneton, the nuclear magneton and the nuclear g-factor, respectively.
In the last step of the above equation, the non-relativistic values of the ionic wavefunctions at the origin
together with the operator relation
\begin{equation}
\hat{\bn S} \hat{\bn I}=\frac{1}{2}\Big( \hat{\bn F}^2 - \hat{\bn I}^2 -\hat{\bn S}^2 \Big)
\label{eq:opereq}
\end{equation}
have been used. Equation \eqref{eq:opereq} is valid as long as the electron states we use are spherically symmetric, that is if they are $s$-states. \\
A property of Eq.\ti(\ref{eq:ev}) to remark is that only states with the same quantum number $F$ can be coupled by the hyperfine interaction Hamiltonian.
For $n=1$ and $m=2$, the mixing coefficient $\mathcal{C}$ in Eq.\ti(\ref{eq:ev}) turns out to be of order $10^{-8}$ in Hydrogen atom while $10^{-5}$ in 
hydrogenlike Uranium ion. These two numbers can be taken as estimation for the accuracy of equation \eqref{eq:totstate1}, for the problem under 
consideration.\\
Finally, due to parity invariance of the hyperfine interaction Hamiltonian,
we underline that states with different parities may not get mixed.
We will discuss
our results in Sect.\ti\ref{sec:results} on the base of the considerations here presented. \\
To further proceed, 
we suppose that the nucleus does not interact with the radiation field. In the language of quantum 
mechanics, this equates to considering that the interaction Hamiltonian couples only electron fields
through photon emission, while it does not act on the quantum space of nuclear states.
This hypothesis holds for decays
which involve ionic bound states, since the energy released
in such decays is far lower than the normal nuclear excitation energies. In contrast,
in cases where the initial ionic state is a (positive) continuum--state,
such hypothesis might fall as the energy of the radiation emitted in consequence of the electron capture might 
be enough to excite the nucleus and, therefore, to give rise to the so--called {\it nuclear excitation by electron capture} (NEEC) \cite{Pa210}. \\
As a result of this supposition, we will find in the next subsection that the radial part of the decay 
amplitude will be characterized by only electron state components.
Whereas, in the angular part, 
both the electron and nucleus states components will play a role, due to the coupling of
the angular momenta.

\subsection{Second order transition amplitude}
\label{sec:second}
\begin{figure}[t]
\centering
\includegraphics[width=.48\textwidth]{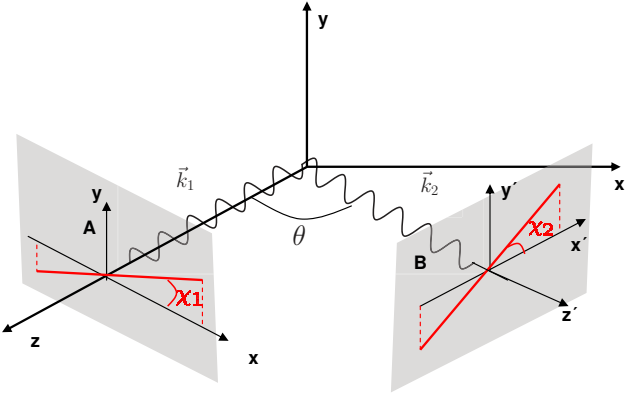}
\caption{(Color online) Geometry considered for the two-photon emission. The propagation direction of the first
photon is adopted as $z$ direction. $x$ is chosen such that $xz$ is the reaction plane 
(plane spanned by the two photon directions). $\theta$ is the angle between the photons directions, while angles $\chi_{1,2}$
define the linear polarizations of the first and second photon respectively with respect to their respective polarization planes. 
The polarization plane of the first (second) photon is denoted by $A$ ($B$) and represents the plane 
orthogonal to the photon direction.
}
\label{fig:fig1}
\end{figure}

The theory of two--photon decay is based on the second order transition amplitude, discussed
for example in Akhiezer and Berestetskii \cite{Ak65}.
For an initial state $\ket{i}$ and a final state $\ket{f}$, such amplitude reads
\begin{equation}
\begin{array}{l}
   \ds\mathcal{M}^{\lambda_1,\lambda_2}(i\to f) = \\
   \quad\ds \ddsum{\nu}\!\!\!\!\!\!\!\!\int
   \Big[
   \frac{ \bra{f} 
   \vec\alpha\cdot\vec u_{\lambda_1}^*e^{-i\vec k_1 \cdot \vec r}
   \ket{\nu}\bra{\nu} 
   \vec\alpha\cdot\vec u_{\lambda_2}^*e^{-i\vec k_2 \cdot \vec r}
   \ket{i}}{E_{\nu}
   -E_i+\omega_2 } \\[0.4cm]
   \qquad\qquad \ds + \, \left( 1 \longleftrightarrow 2\right) \Big] ~ ,
   \end{array}
   \label{Mfi}
\end{equation}
where $\vec k_{j}$, $\vec u_{\lambda_j}$, $\lambda_{j}$ and $\omega_j$ are respectively the wave vector,
the polarization vector, the helicity and the energy of the $j$th emitted photon ($j=1,2$). $E_{i,\nu}$ are the energies of
the initial and intermediate ionic state, while $\vec \alpha$ is
the standard vector of Dirac matrices.
The subscripts $i$, $\nu$, $f$ will be used throughout the paper as referring to initial, intermediate and final state respectively.
The summation over the intermediate states 
showed in formula (\ref{Mfi})  runs over the whole ionic spectrum, including 
a summation over the discrete part as well as an integration over the (positive and negative) continuum.
For the problem under consideration, such summation splits up into summations
over the principal quantum number $n_{\nu}$, the Dirac quantum number $\kappa_{\nu}$, the total angular momentum $F_\nu$ and
its projection onto the quantization axis $m_{F_{\nu}}$. \\
Since the two emitted photons define generally two different trajectories, the evaluation of the angular properties of the 
amplitude in Eq.~(\ref{Mfi}) is best carried out by decomposing
the photon fields into their spherical tensor components of defined angular momentum. For the $j$th photon, such
a decomposition reads \cite{Ro53}
\begin{equation}
\begin{array}{l c l}
\vec u_{\lambda_{j}}e^{i\vec k_{j}\cdot\vec r}&=&\sqrt{2\pi}\,{\ds \sum_{L=1}^{+\infty}\sum_{M=-L}^{L}}i^L[L]^{1/2}\left(
\vec A_{L\,M}^{(m)} \right.\\[0.6cm]
&&\left. +i\lambda_{j}\vec A_{L\,M}^{(e)}\right)
  D_{M\,\lambda_{j}}^L(\hat k_{j}\to \hat z) ~,
\end{array}
\label{eq:phdecomp}
\end{equation}
where $[L]=2L+1$ while $D_{M\,\lambda_{j}}^L(\hat k_{j}\to \hat z)$
are the Wigner rotation matrices which rotate each tensor component with angular momentum $L$ and original quantization axis $\hat k_{j}$ into
the same component with the chosen quantization axis $\hat z$.
Furthermore, the standard notation $\vec A_{L\,M}^{(e)}$ and $\vec A_{L\,M}^{(m)}$ is 
used to denote the electric and magnetic multipole, respectively.
Each one of these multipoles can be expressed in terms of the spherical Bessel functions $j_L(kr)$ and the spherical
tensor $\vec T_{L\,\Lambda\,M}$ as 
\begin{equation}
\begin{array}{l c l}
\vec A_{L\,M}^{(m)} &=&  j_L(kr) \vec T_{L\,L\,M}(\hat r) \\
\vec A_{L\,M}^{(e)} &=&  j_{L-1}(kr) \sqrt{\frac{L+1}{2L+1}}\vec T_{L\,L-1\,M}(\hat r)\\
&& -j_{L+1}(kr) \sqrt{\frac{L}{2L+1}}\vec T_{L\,L+1\,M}(\hat r) ~. \\
&&
\end{array}
\label{eq:sphericalcom}
\end{equation}
Upon introducing Eq.~(\ref{eq:phdecomp}) into (\ref{Mfi}), expanding the ionic states
as showed in Eq.~(\ref{eq:totstate1}) and by taking into account that 
the nuclear states must be normalized, the amplitude can be written as
\begin{widetext}
\begin{equation}
\begin{array}{c}
\ds\mathcal{M}^{\lambda_1,\lambda_2}(i\to f) =
-(2\pi)\ddsum{T\,T'}\;\ddsum{\substack{\kappa_{\nu}\\m_I\,m_{j_{\nu}}}}\;
\ddsum{\substack{L_1\,L_2\\M_1\,M_2}}\;\ddsum{p_1\,p_2}\;\ddsum{\Lambda_1\,\Lambda_2}
(\lambda_1)^{p_1}(\lambda_2)^{p_2}[L_1,L_2]^{1/2}
i^{-L_1-L_2-p_1-p_2}\, \xi_{L_1\,\Lambda_1}^{p_1}
\xi_{L_2\,\Lambda_2}^{p_2} P^T \; P^{T'} \\[0.4cm]
\times\;\ds D^{L_2\,*}_{M_2\,\lambda_2}(\hat k_2\to\hat z)D^{L_1\,*}_{M_1\,\lambda_1}(\hat k_1\to\hat z) 
\Bigg[ 
U^{TT'}_{\Lambda_1\,\Lambda_2} \; \chi_{m_I\,m_{j_{\nu}}}^{f^T\, \nu^T} \chi_{m_I\,m_{j_{\nu}}}^{\nu^{T'}\, i^{T'}} 
\;+\; \big(1\leftrightarrow 2\big) \Bigg] ~,
\end{array} 
\label{Mfi2}
\end{equation}
\end{widetext}
where $\Lambda_{j}$ runs from $L_{j}-1$ to $L_{j}+1$, $T=L,S$ is used to denote the large (L) and small (S) components of the electron Dirac spinor, 
for which the factor $P^T$ is defined as $P^L=1$ and $P^S=-1$,
where $\bar T$ refers to the conjugate of $T$, $i.e.$ $\bar T=L$ for $T=S$ and vice versa. Furthermore,
$p_{1,2}=0,1$ and the function $\xi_{L\,\Lambda}^p$ is given by
\begin{equation}
\begin{array}{l c l}
\xi_{L\,\Lambda}^0&=&\delta_{L,\,\Lambda} \qquad,\\[0.2cm]
\xi_{L\,\Lambda}^1&=&\left\{
\begin{array}{l r}
\sqrt{\frac{L+1}{2L+1}} & \textrm{for } \Lambda=L-1\\[0.2cm]
-\sqrt{\frac{L}{2L+1}} & \textrm{for } \Lambda=L+1\\[0.2cm]
0 & \textrm{otherwise } ~.
\end{array}
\right.
\end{array}  
\label{eq:csi}
\end{equation}
The radial part of the amplitude in Eq.~(\ref{Mfi2}) is represented by the integral $U^{TT'}_{\Lambda_1\,\Lambda_2}$, which reads
\begin{equation}
U^{TT'}_{\Lambda_1\,\Lambda_2}=
\int \dd r \dd r' r^2r'^2 j_{\Lambda_1}(k_1r')j_{\Lambda_2}(k_2r) 
g_f^{\bar{T}*} g_{E_i+\omega_1}^{T\,\bar{T'}} g_i^{T'} ~,
\label{eq:int}
\end{equation}
where $g_{f,i}^T$ are the small and large radial components of the final and initial electron state,
while  
\begin{equation}
g_{E_i+\omega_1}^{T\,\bar{T'}}=\ddsum{n_{\nu}}\frac{g_{\nu}^T\,g_{\nu}^{\bar{T'}*}}{E_{\nu}-E_i-\omega_1}
\label{eq:green}
\end{equation}
is the radial Green function of the process.\\
By inspecting Eq.~(\ref{eq:int}), we notice that the radial part of the transition amplitude involves only electron state components.
As mentioned in Sect.~\ref{sec:states}, this fact comes directly from having neglected any photon--nucleus interaction.
Although the integral in Eq.~(\ref{eq:int}) involves only electron state components, its evaluation is anyway a challenging task
due to the (infinite) summation over the principal quantum number $n_{\nu}$ contained in the radial Green function.
In the present work, such integral has been computed by using the Greens library \cite{Ko03}.
This approach has been presented for the first time
in the context of two--photon decays
by Surzhykov {\it et al.} in Ref.\cite{An05}. \\
The angular part of the amplitude in Eq.\ti(\ref{Mfi2}) is represented by
the elements $\chi_{m_I\,m_{j_{\nu}}}^{f^T\, \nu^T}$ and $\chi_{m_I\,m_{j_{\nu}}}^{\nu^{T'}\, i^{T'}}$ therein contained
and can be computed analytically.
With the help of Eq.~(\ref{eq:totstate1}), we can indeed write
\begin{equation}
\begin{array}{l}
\chi_{m_I\,m_{j_{\nu}}}^{f^T\, \nu^T} = \ddsum{m_{j_f}}\braket{j_f,m_{j_f},I,m_I}{F_f,m_{F_f}} \\
\qquad\times\;
\bra{\kappa_f, l_f^T, m_{j_f}} \vec\sigma\cdot\vec T^*_{L_2\,\Lambda_2\,M_2}\ket{\kappa_{\nu}, l_{\nu}^{\bar T}, m_{j_{\nu}}}\\[0.4cm]
\chi_{m_I\,m_{j_{\nu}}}^{\nu^{T'}\, i^{T'}} = \ddsum{m_{j_i}}\braket{j_i, m_{j_i}, I, m_I}{F_i, m_{F_i}} \\
\qquad \times\;
\bra{\kappa_{\nu}, l_{\nu}^{T'}, m_{j_{\nu}}}\vec\sigma\cdot\vec T_{L_1\,\Lambda_1\,M_1}^*\ket{\kappa_i, l_i^{\bar T'}, m_{j_i}} ~,
\end{array}
\label{eq:angpart}
\end{equation}
where $\vec\sigma$ are Pauli matrices while
the elements $\bra{\kappa_f,l_f^{T},m_{j_f}}$$\vec\sigma\cdot\vec T_{L_2\,\Lambda_2\,M_2}^* $$\ket{\kappa_{\nu},l_{\nu}^{\bar{T}},m_{j_{\nu}}}$
and $\bra{\kappa_{\nu}, l_{\nu}^{T'}, m_{j_{\nu}}}$$\vec\sigma\cdot\vec T_{L_1\,\Lambda_1\,M_1}^*$$\ket{\kappa_i, l_i^{\bar T'}, m_{j_i}}$
have been already discussed elsewhere \cite{An05, An02} and will not be here recalled. \\
The initial and final states involved in the two--photon transitions which we shall analyze in Sect.~\ref{sec:results} are $unpolarized$.
Therefore, since for the decay of unpolarized states there is not any preferred direction for the overall system,
we arbitrarily adopt the quantization axis ($\hat z$) along the momentum of the \supercomas{first} photon: $\hat z\parallel \hat{k}_1$.
We furthermore adopt $\hat x$ such that the $xz$-plane is the reaction plane (plane spanned by the photons directions).
Figure \ref{fig:fig1} sketches the geometry we consider for the decay.
Within this geometry,  
the Wigner matrices in Eq.~(\ref{Mfi2}) simplify to 
\begin{equation}
\begin{array}{l}
D^{L_1\,*}_{M_1\,\lambda_1}(\hat k_1\to\hat z)=\delta_{M_1,\,\lambda_1} ~,\\[0.4cm]
D^{L_2\,*}_{M_2\,\lambda_2}(\hat k_2\to\hat z)=d^{L_2}_{M_2\,\lambda_2}(\theta) ~,
\end{array}
\end{equation}
where $d^{L}_{M\,\lambda}(\theta)$ is the reduced Wigner matrix and $\theta$ is the polar angle of the second photon, which coincides, in the 
chosen geometry, with the angle between the photons directions (opening angle). Hence, the relative photons directions
are uniquely determined by assigning the opening angle $\theta$, which will be the variable against which the angular correlation
will be plotted in Sect.~\ref{sec:results}.

\subsection{Angular and Polarization correlations}
\label{sec:ddr}
Within the approximations mentioned in Sect.~\ref{sec:states}, equation (\ref{Mfi2}) represents the relativistic
transition amplitude for the two--photon decay between hyperfine states in hydrogenlike ions.
It contains the complete information on the emitted radiation.
For instance, the energy and angle-polarization distributions of the two emitted photons
can be 
written in terms of such transition amplitude.
Assuming that the ion is initially unpolarized and that the polarization
of the final ionic state remain unobserved, the differential
decay rate reads (atomic units) \cite{Go81}
\begin{equation}
\begin{array}{l}
\ds \frac{\dd w^{\lambda_1 \lambda_2}}{\dd \omega_1 \dd \Omega_1 \dd \Omega_2}=
\frac{\omega_1\omega_2}{(2\pi)^3c^2}\frac{1}{2F_i+1}\\[0.4cm]
\qquad\times\;\ddsum{m_{F_i}\,m_{F_f}}
\Big|\mathcal{M}^{\lambda_1\lambda_2}(i\to f)\Big|^2 ~.
\end{array}
\label{eq:ddr}
\end{equation}
By taking into account the axes geometry chosen for the two--photon emission,
which has been explained in Sect.~\ref{sec:second}, we can easily perform 
the integration over the solid angle of the first photon ($\dd \Omega_1$) as well as 
the integration over the azimuthal angle of the second photon ($\dd \phi_2$). Now, since part of this work is devoted
to analyze photons polarizations, further details concerning the detection geometry must be provided, in order to proceed with the analysis\\
In Fig.~\ref{fig:fig1}, 
we show how the photon polarizations may be defined in a case experiment.
The polarization of each photon is to be measured in the \supercomas{polarization plane}, which is
the plane orthogonal to the photon direction. 
This means that, without loss of generality, we are considering linear (not circular) polarizations of photons.
This choice is motivated by the fact that, in high energy regime, linear polarizations of photons are much more easily measured than 
circular polarizations.
In Fig.~\ref{fig:fig1}, 
the polarization plane of the first and second photon are denoted by $A$ and $B$, respectively.
Each detector is 
supposed to have a transmission axis, along which the linear polarization of the photon is measured. Such a transmission axis is rotated
by an angle $\chi$ with respect to the reaction plane. Finally, each detector is supposed to work as a filter: 
Whenever a photon hits it, the detector gives off or not a \supercomas{click}, which would respectively
indicate that the photon has been measured as having its linear polarization along the direction $\chi$ or $\chi+90^{\circ}$. \\
By integrating over the first photon energy ($\dd \omega_1$) and by using the well-known relations between linear and circular polarization bases \cite{Ro53}
\begin{equation}
   \vec u_{\chi} = 
   \frac{1}{\sqrt{2}} \, \left( {\rm e}^{-i\chi} \vec u_{\lambda=+1} +
   {\rm e}^{+i\chi} \vec u_{\lambda=-1} \right) \, ,
   \label{eq:circTOlin}
\end{equation}
we can write the
differential decay rate which depends upon the photons linear polarizations as
\begin{equation}
\begin{array}{l}
\ds W^{\chi_1\,\chi_2}(\theta)\equiv \frac{\dd w^{\chi_1 \chi_2}}{\dd \cos\theta }=
\frac{8\pi^2}{2F_i+1}
\ddsum{m_{F_i}\,m_{F_f}}\ddsum{\substack{\lambda_1\lambda_2\\\lambda_1'\lambda_2'}}\\[0.6cm]
\ds\qquad\times\int \dd\omega_1 \;\frac{\omega_1\omega_2}{4(2\pi)^3c^2}\; e^{i(\lambda_1-\lambda_1')\chi_1}
e^{i(\lambda_2-\lambda_2')\chi_2}\\[0.6cm]
\ds\qquad\times \;\mathcal{M}^{\lambda_1\lambda_2}
\mathcal{M}^{\lambda_1'\lambda_2'\,*}~.
\end{array}
\label{eq:ddrLP}
\end{equation}
We shall call this function \supercomas{polarization correlation}. The polarization correlation represents 
the probability of detecting the emitted photons with defined linear polarizations $\chi_1$ and $\chi_2$, 
in the two--photon decays of hyperfine states in hydrogenlike ions.\\
Finally, by summing over the photons polarizations, we define the \supercomas{angular correlation} function as
\begin{equation}
\begin{array}{l}
\ds W(\theta)\equiv\frac{\dd w}{\dd \cos\theta}=
\frac{8\pi^2}{2F_i+1}\int \dd \omega_1 \;
\frac{\omega_1\omega_2}{(2\pi)^3c^2}\\[0.4cm]
\qquad\times\;\ds\ddsum{m_{F_i}\,m_{F_f}}\ddsum{\lambda_1\lambda_2}
\Big|\mathcal{M}^{\lambda_1\lambda_2}(i\to f)\Big|^2 ~.
\end{array}
\label{eq:W}
\end{equation}
Equipped with equations \eqref{eq:ddrLP} and \eqref{eq:W},
in the next
section we shall investigate the functions $W^{\chi_1\,\chi_2}(\theta)$ and $W(\theta)$ as obtained for a few decays
of unpolarized hyperfine $2s$ states in Hydrogen atom and in hydrogenlike Uranium ion.
\begin{figure}
\centering
\includegraphics[width=.48\textwidth]{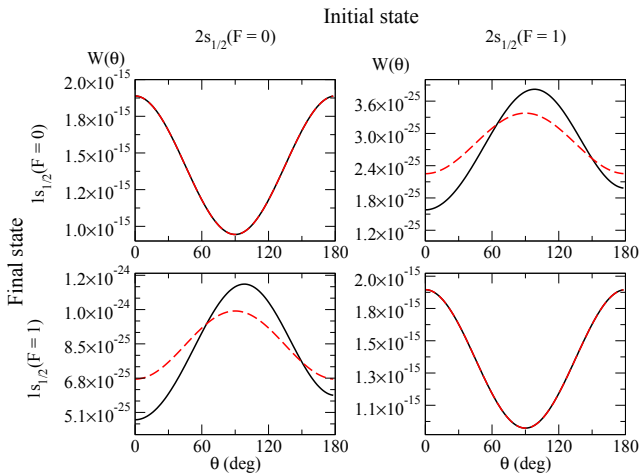}
\caption{(Color online) Angular correlation $W(\theta)$ for the transitions 
$2s_{1/2}\,(F=1,0)\to1s_{1/2}\,(F=1,0)$ in Hydrogen atom. 
The dashed-red curve refers to the electric dipole approximation while
the solid-black curve refers to
the full multipole contribution. 
}
\label{fig:fig2}
\end{figure}
%
%
%
%
%
%
%
%
%
%
%
%
%
%
%
\section{Results and discussion}
\label{sec:results}
In this section we shall analyze the 
angular and polarization correlations defined in Eqs.~\eqref{eq:ddrLP} and \eqref{eq:W}, for decays of hyperfine $2s$ states in
Hydrogen atom as well as in hydrogenlike Uranium ion. 
\subsection{Angular and polarization correlations in Hydrogen atom}
\begin{figure}[t]
\centering
\includegraphics[width=.48\textwidth]{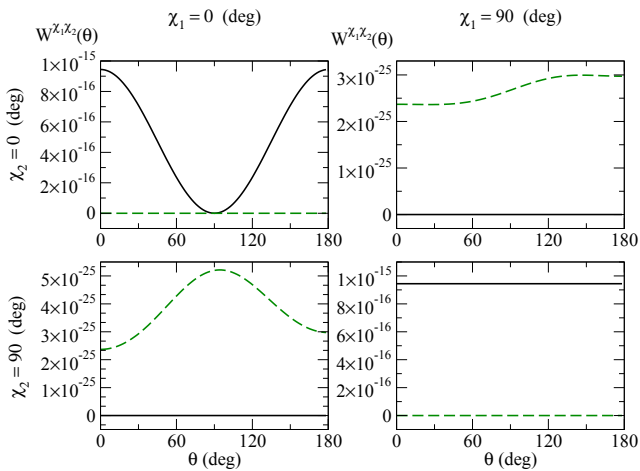}
\caption{(Color online) Polarization correlation $W^{\chi_1\,\chi_2}(\theta)$ for the transitions
$2s_{1/2}\,(F=0)\to1s_{1/2}\,(F=0)$ (solid-black curve) and $2s_{1/2}\,(F=0)\to1s_{1/2}\,(F=1)$ (dashed-green curve) in Hydrogen atom.
The four polarization configurations ($\chi_1, \chi_2=0^{\circ}, 90^{\circ}$) are separately displayed.
}
\label{fig:fig3}
\end{figure}

We first analyze Hydrogen atom, whose nuclear spin is, as known, I = 1/2. In particular, we analyze
the four two--photon transitions
$2s_{1/2}\,(F=1,0)\to1s_{1/2}\,(F=1,0)$. 
The predicted angular correlation $W(\theta)$ for these transitions is displayed in Fig.~\ref{fig:fig2}, where the full
multipole and the electric dipole (E1E1) contributions are separately displayed.
We might immediately notice that, in the two transitions $2s_{1/2}\,(F=1)\to1s_{1/2}\,(F=1)$
and $2s_{1/2}\,(F=0)\to1s_{1/2}\,(F=0)$, the curves obtained within the electric
dipole approximation practically coincide with the curves obtained with the full multipole contribution. The angular correlation
in these transitions can be
well described by $W(\theta)\sim1+\cos^2\theta$, which is {\it symmetric} 
with respect to $\theta=90^{\circ}$
and which is the 
analytical form that has been for long associated to the angular correlation 
in $2s\to1s$ two--photon transitions in Hydrogenlike bound systems \cite{MoD04, Au76, Kl69}. \\
Whereas, the radiation patterns
in $2s_{1/2}\,(F=1)\to1s_{1/2}\,(F=0)$ and $2s_{1/2}\,(F=0)\to1s_{1/2}\,(F=1)$ transitions show a different, {\it asymmetric} shape.
It can be seen that high-multipole contributions
are directly responsible for the asymmetry,
while the contribution of the leading E1E1 multipole  
can be well described by $W_{E1E1}(\theta)\sim1-1/3\cos^2\theta$.
This result is typical for two-photon
decays of the type J$_{TOT}$ = 1 (0) $\to$ J$_{TOT}$ = 0 (1), where J$_{TOT}$ is the total angular momentum of the system
which undergoes the decay. In fact the two-photon decay $(1s\,2s)^3S_{J=1} \to
(1s\,1s)^1S_{J=0}$ in heliumlike ions, where $J$ is the total angular momentum of the system given by the coupling of the two electron
angular momenta, shows approximately the same behavior \cite{An10, PS69}.\\
Quantitatively, the asymmetry 
\begin{equation}
\mathcal{A}=\frac{W(\theta=180^{\circ})-W(\theta=0^{\circ})}{W(\theta=180^{\circ})}
\label{asy}
\end{equation} 
is approximately 0.20 in F = 1 (0) $\to$ F = 0 (1) transitions, while it is vanishing in F = 1 (0) $\to$ F = 1 (0) transitions.\\
Difficulties in measuring the correlation function in those transitions where the total spin ($F$) is flipped 
(F = 0 (1) $\to$ F = 1 (0)) might arise from the fact that the decay rates for these transitions 
are noticeably nine orders of magnitude smaller than the decay rates for the transitions in which the spin is not flipped
(F = 0 (1) $\to$ F = 0 (1)), as seen from Fig.~\ref{fig:fig2}.
This fact is caused by the strong cancellation of the contributions given by the $p_{1/2}$ and $p_{3/2}$ intermediate
states (in the calculation of the leading multipole E1E1) to the correlation function in F = 0 (1) $\to$ F = 1 (0) transitions. 
As a matter of fact, within non-relativistic dipole approximation, these two contributions 
are equal and opposite,
so that the transitions $2s_{1/2}\,(F=1)\to1s_{1/2}\,(F=0)$ and $2s_{1/2}\,(F=0)\to1s_{1/2}\,(F=1)$
cannot proceed via two-photon dipole E1E1 emission \cite{An10, Ke83}.
Thus, the non-vanishing correlation functions showed in the top-right and bottom-left panels of Fig.~\ref{fig:fig2} are only given by
high-multipole and relativistic effects.\\
As a consequence of what said above, any experimental incoherent population of the 
initial and final hyperfine states would result in measuring the angular correlation function as given by
$\sim1+\cos^2\theta$. This has been confirmed, for example, in Refs.~\cite{Li65, OCo75}. 
For a direct experimental investigation of the angular correlation function in
F = 1 (0) $\to$ F = 0 (1) transitions,
one would therefore
normally need to selectively populate the initial and observe the final hyperfine state.
Such a selection could be made, in principle, by resolving energetically the emitted photons. The two hyperfine states
$2s\,(F=1)$ and $2s\,(F=0)$, as well as the states $1s\,(F=1)$ and $1s\,(F=0)$, although
they have been considered degenerate in the present paper, possess in fact slightly different energies due to the well-known hyperfine 
energy splitting, which is of the order $\sim10^{-6}$ eV in both cases (see Sect.\ti\ref{sec:states}). 
However, the energy resolution of normal photon detectors for light quanta in the visible range,
which is the energy range of the photons emitted in the considered transitions, is far lower than $10^{-6}$ eV.
Therefore, the selection of initial and final hyperfine states cannot proceed by analyzing energy.\\
Alternatively, in order to achieve the same goal,
we could make use of the linear polarization properties of the emitted photons. 
In Fig.~\ref{fig:fig3}, we plot the polarization correlation $W^{\chi_1\,\chi_2}(\theta)$ as obtained for the transitions $2s_{1/2}(F=0)\to 1s_{1/2}(F=0)$ and 
$2s_{1/2}(F=0)\to 1s_{1/2}(F=1)$ in Hydrogen atom. 
As it can be seen, the orthogonal linear polarization configurations $\chi_1=0^{\circ}, \chi_2=90^{\circ}$ and $\chi_1=90^{\circ}, \chi_2=0^{\circ}$
are absent in F = 0 $\to$ F = 0 transition.
More quantitatively, from Fig.~\ref{fig:fig3} we can identify the photons' polarization state for F = 0 $\to$ F = 0 transition as being
\begin{equation}
\begin{array}{l c l}
\ket{\Psi}&=&\ds\frac{1}{\sqrt{1-\cos^2\theta}}\Big( \ket{\chi_1=0, \chi_2=0} \\[0.3cm]
&&\qquad+ \,e^{i\delta}\cos\theta\ket{\chi_1=90^{\circ}, \chi_2=90^{\circ}} \Big) ~,
\end{array}
\label{state}
\end{equation}
where the phase $\delta$ can be determined, for instance, by analyzing the polarization correlation in the helicity basis.
By doing that, one easily finds $\delta=0$.
Although not directly shown, 
the photons polarization state in Eq.~(\ref{state}) holds also in case the photons came from F = 1 $\to$ F = 1 transition.\\
In Refs.~\cite{PolPol, QuantInf}, 
the same photons polarization state has been derived by analyzing $2s\to1s$ transitions 
in hydrogenlike ions within non-relativistic dipole approximation and without considering the nuclear spin.
This is consistent with what found above, 
since, as mentioned, F = 0 (1) $\to$ F = 1 (0) transitions are forbidden within non-relativistic dipole approximation and therefore
the polarization state displayed in Eq.~(\ref{state}) is the only one which the photons can assume in two-photon $2s\to1s$
transitions.\\
Coming back to Fig.~\ref{fig:fig3}, we notice that 
the decay rate for orthogonal polarization configurations is not vanishing 
for F = 0 $\to$ F = 1 transition, for any angle $\theta$, although its order of 
magnitude is very small in comparison with the decay rates shown in other panels of the same figure.
This fact allows us to suggest a workable scenario for investigating the angular and polarization
correlations in two--photon F = 0 $\to$ F = 1 transitions.
By using standard devices of modern spectroscopy, we can experimentally exclusively populate 
the initial $2s_{1/2}\,(F=0)$ state \cite{Ro00}. Such a state would consequently decay either into 
$1s_{1/2}\,(F=0)$ or $1s_{1/2}\,(F=1)$ state. 
Now, as already underlined,
only the transition $2s_{1/2}\,(F=0)\to 1s_{1/2}\,(F=1)$ of the two is characterized by non-vanishing decay rate for two photons
having orthogonal linear polarizations. Thus, we could select out the two-photon decay channel F = 0 $\to$ F = 1
just by filtering the photons polarizations.
By virtue of the fact that the wavelength of each emitted quanta in this decay
is roughly $\approx 250$ nm and that polarizer crystals are available in such wavelength range, we could place 
one linear polarizer filter in front of each photon detector. We would use the detectors in
coincidence and we would set opposite linear polarizers transmission axes of the filters,
in order to allow only the photon pairs coming from $2s_{1/2}\,(F=0)\to 1s_{1/2}\,(F=1)$ transition to be detected.
By inspecting the orders of magnitude of the decay rates in Fig.~\ref{fig:fig3},
we notice that, in order to successfully block the unwanted light from F = 0 $\to$ F = 0 transition, 
the extinction ratio of the polarizers must be higher than 10$^{10}$:1. Such performances should be achievable by 
using in series, for instance, two or more normal polarizers whose extinction ratio is usually of order 10$^{5}$:1.
Problems related to the alignment of the polarizers might however arise.\\
By using this sketched set-up, we should therefore 
be sensible only to those photons that come from $2s_{1/2}\,(F=0)\to 1s_{1/2}\,(F=1)$ transition and that have
the selected polarization configuration, $i.e.$ we will measure the dashed-green curve showed either in the bottom-left or
in the top-right panel of Fig.~\ref{fig:fig3}, depending on the polarizers settings.\\
Due to the intrinsic weakness of the signal coming from the desired decay channel,
one could at this point think that the hyperfine interaction itself, here neglected from the outset,
may mix states with different quantum numbers, so that the angular
and polarization properties of the emitted radiation could vary from what shown in Figs.~\ref{fig:fig2} and \ref{fig:fig3}. 
However, as briefly discussed in Sect.\ti\ref{sec:states}, the Hamiltonian which accounts for the hyperfine interaction
neither mixes states with different parities nor $S$-states with different quantum number $F$. This implies that 
the state $2s_{1/2}\,(F=0)$ does not acquire any admixture of any other state and thus remains well defined.
Indeed low energy hyperfine states in Hydrogen atom are known to be well separated, as it is evident from the fact that the light
emitted in the hyperfine transition of $1s_{1/2}$ state in Hydrogen is one of the most accurately measured physical quantities ever and is of
great importance in astrophysics and cosmology.
\begin{figure}
\centering
\includegraphics[width=.48\textwidth]{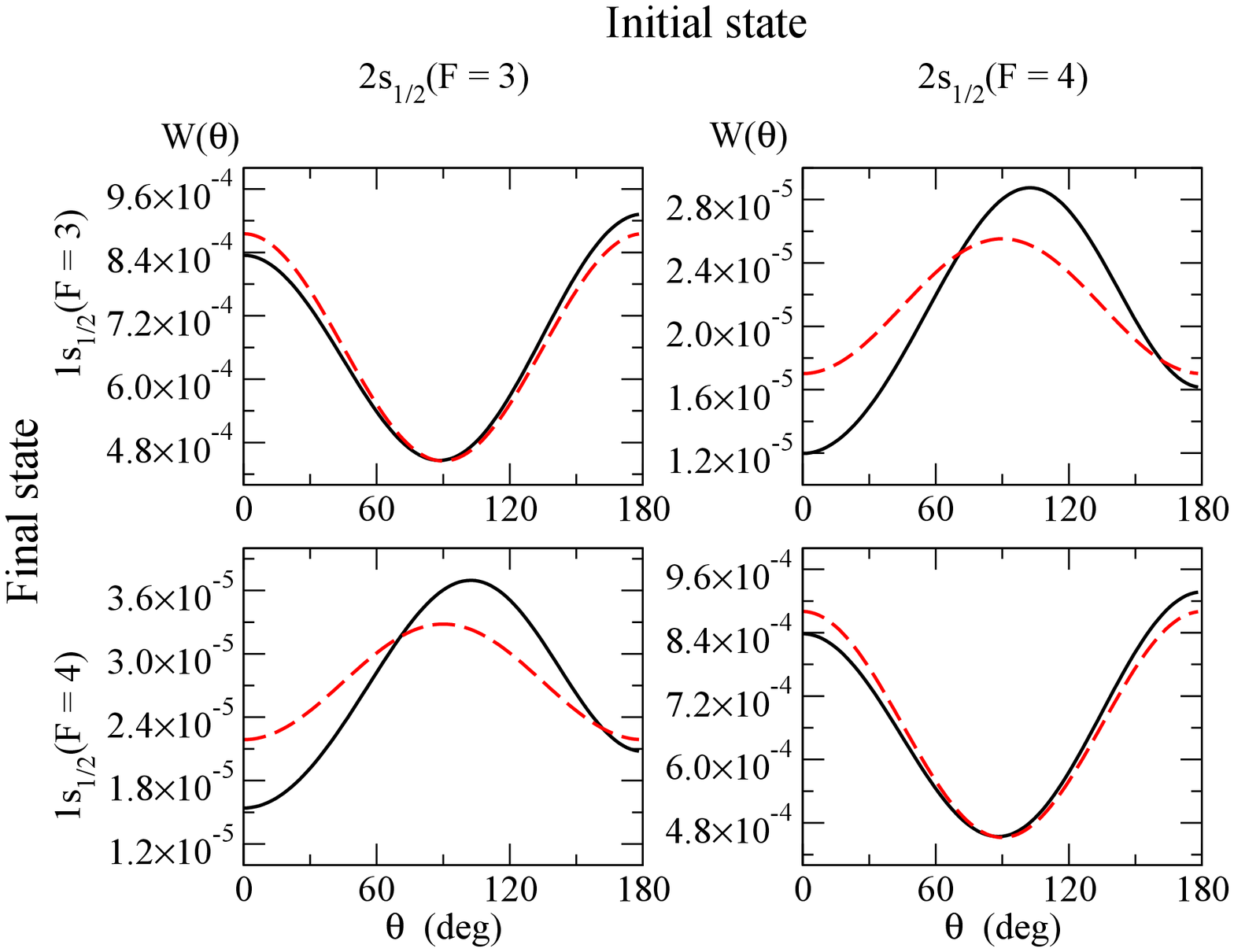}
\caption{(Color online) Angular correlation $W(\theta)$ for the transitions 
$2s_{1/2}\,(F=4,3)\to1s_{1/2}\,(F=4,3)$ in hydrogenlike ${}^{235}_{92}$U. 
The dashed-red curve refers to the electric dipole approximation while
the solid-black curve refers to
the full multipole contribution. 
}
\label{fig:fig4}
\end{figure}
\subsection{Angular and polarization correlations in hydrogenlike Uranium}
\begin{figure}[t]
\centering
\includegraphics[width=.48\textwidth]{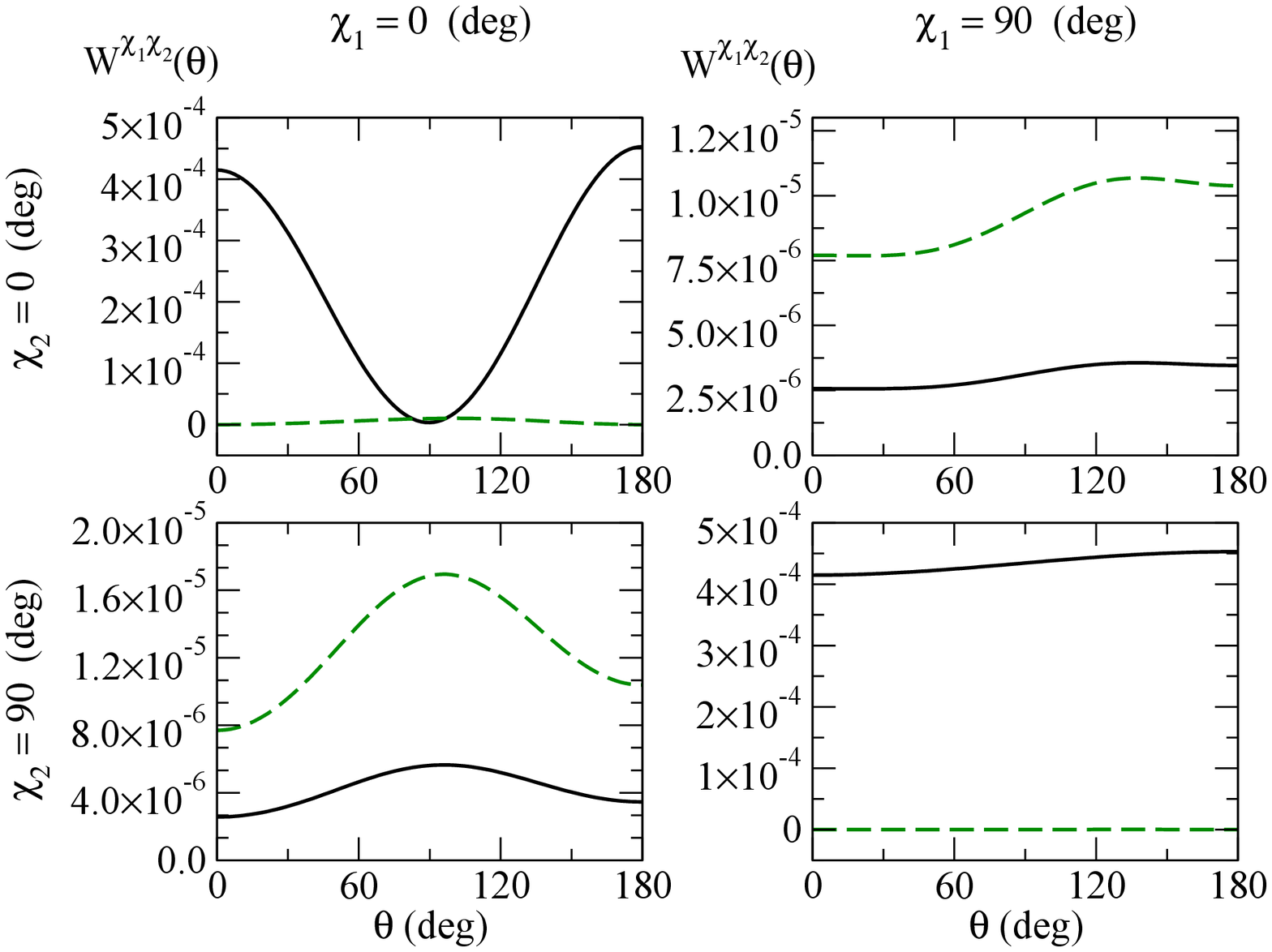}
\caption{(Color online) Polarization correlation $W^{\chi_1\,\chi_2}(\theta)$ for the transitions
$2s_{1/2}\,(F=3)\to1s_{1/2}\,(F=3)$ (solid-black curve) and $2s_{1/2}\,(F=3)\to1s_{1/2}\,(F=4)$ (dashed-green curve) in 
hydrogenlike ${}^{235}_{92}$U ion.
The four polarization configurations ($\chi_1, \chi_2=0^{\circ}, 90^{\circ}$) are separately displayed.
}
\label{fig:fig5}
\end{figure}

We proceed to analyze hydrogenlike ${}^{235}_{92}$U ion, whose nuclear spin is I = 7/2 \cite{St05}.
In particular, we study the transitions $2s_{1/2}\,(F=4,3)\to1s_{1/2}\,(F=4,3)$ in such ion.
Hydrogenlike heavy ions
are interesting bound systems both from a theoretical and an experimental point of view, 
since they are as simple as Hydrogen atom but show, at the same time,
remarkable relativistic and high-multipole effects.
We have nonetheless seen that high-multipole effects show already up in F = 0 (1) $\to$ F = 1 (0) transitions in Hydrogen atom,
where they are directly responsible for the asymmetric behaviour of the angular correlation function (cfr. Fig.~\ref{fig:fig2}).
In ${}^{235}_{92}$U ion, we expect an enhancement of
this effect.\\
The angular correlation, $W(\theta)$, of the photon pair for $2s_{1/2}\,(F=4,3)\to1s_{1/2}\,(F=4,3)$ two--photon transitions 
in hydrogenlike ${}^{235}_{92}$U ion is displayed in Fig.~\ref{fig:fig4}.
The full multipole and the electric dipole (E1E1) contributions are separately displayed.
Although the spin properties of ${}^{235}_{92}$U and ${}^{1}_{1}$H are considerably different, we can see that the relative (photon) angular
correlations are rather similar in shape. 
However, by inspecting carefully both Figs.~\ref{fig:fig4} and \ref{fig:fig2}, it can be noticed that, 
as expected, high-multipole effects play overall a more important role in ${}^{235}_{92}$U than in Hydrogen.
First, in contrast to Hydrogen,
the angular correlation for non-spin-flip transitions (F = 3 (4) $\to$ F = 3 (4)) show sizable deviations from the $\sim1+\cos^2\theta$ shape.
This effect is already known from the past literature, where it has been showed that high multipoles
contribute with terms of the type
$\sim\cos\theta$ to the angular correlation in $2s_{1/2}\to1s_{1/2}$ transitions \cite{Au76}.
Secondly, we also notice that, in spin-flip transitions (F = 3 (4) $\to$ F = 4 (3)),
deviations of the angular correlation function from the symmetric analytical formula $\sim1-1/3\cos^2\theta$ are
more pronounced in hydrogenlike ${}^{235}_{92}$U ion than in Hydrogen. 
More concretely, for the case of ${}^{235}_{92}$U ion, the asymmetry $\mathcal{A}$ defined in Eq.~(\ref{asy}) turns out to
be approximately 0.26 in $2s_{1/2}\,(F=4(3))\to1s_{1/2}\,(F=3(4))$ transitions, 
while it is 0.09 in $2s_{1/2}\,(F=4(3))\to1s_{1/2}\,(F=4(3))$ transitions.\\
It can be furthermore noticed that
each one of the $2s\to1s$ two--photon transitions shown in Fig.~\ref{fig:fig4} is characterized by far higher decay rate with
respect to the transitions shown in Fig.~\ref{fig:fig2}. 
This fact is not surprising since it is known that
the total decay rate for this transition rises fast with $Z$, as it 
can be approximately expressed by $\sigma\approx8.226$ Z$^6$ sec$^{-1}$ \cite{Sh58}.\\
As a final remark, by comparing Figs.~\ref{fig:fig4} and \ref{fig:fig2}, we notice 
that spin-flip transitions are less suppressed in hydrogenlike Uranium than in Hydrogen atom, which is a consequence of the fact that
relativistic and retardation effects characterize highly charged ions.%
The selective detection of the photon pair coming from F = 3 (4) $\to$ F = 4 (3) 
transitions in hydrogenlike Uranium, however, is complicated by the fact that the energy of the emitted radiation 
is in the range of hard X-rays.
The discernment by energy analysis of the decay channels in ${}^{235}_{92}$U 
share the same problem already encountered in the Hydrogen 
case: 
The hyperfine energy splitting between 
$2s_{1/2}\,(F=4)$ and $2s_{1/2}\,(F=3)$, as well as the one between $1s_{1/2}\,(F=4)$ and $1s_{1/2}\,(F=3)$,
is of the order $\sim$ 1 eV, which is far smaller than the normal 
resolution of X-ray detectors. 
In addition, since there exist no polarizer filters for such energy regime,
the photon polarization analysis suggested to select the desired decay channel in the Hydrogen case 
is not immediately applicable to Uranium, regardless of the polarization properties of the emitted photons.
As a matter of fact, the polarization resolved experiments in X-ray energy regimes are not normally
performed by using polarizer filters, but rather by using Compton polarimeters \cite{TNS52a, TNS52b, RSI79, JI5, RSI67}. 
In such devices, the information about the photon polarization is obtained
from the reconstruction of the Compton scattering events (of the incident photons) 
in the detector and thus has a statistical nature. 
As a consequence, Compton polarimeters are not normally used to record the polarization properties in multi-photon processes. Rather,
they are employed to record the polarization properties in single photon processes \cite{PRL97}. Nonetheless,
by selecting events which have been recorded in coincidence by two or more Compton polarimeters and which have the desired scattering angle,
information on the polarization of a multi-photon state could be in principle 
achieved \cite{PNC, NatWard, PRSny, PRWu, PRBleu}. \\
In the light of such possibility, 
we present in Fig.~\ref{fig:fig5} the function $W^{\chi_1\chi_2}(\theta)$
as obtained for the transitions $2s_{1/2}(F=3)\to 1s_{1/2}(F=3)$ and 
$2s_{1/2}(F=3)\to 1s_{1/2}(F=4)$ in hydrogenlike ${}^{235}_{92}$U ion. We see from the figure that
the general scheme for which photons coming from spin-flip transitions (non-splin flip transitions) have
mainly orthogonal (parallel) linear polarizations remains valid also in hydrogenlike Uranium. 
However, in the Uranium case, 
the suppressed polarization configurations are more evident, especially in spin-flip transitions, where they turn out to be of the same order
of magnitude of the leading polarization configurations.
Thus, in contrast to
Hydrogen, there exist not any 
polarization configuration by which to single out the photons coming from one transition of the two
displayed in Fig.~\ref{fig:fig5}.
\section{Summary}
\label{sec:summary}
In summary, the angular and polarization correlations in two--photon decays of $2s$ hyperfine states in hydrogenlike ions have been investigated 
within second-order perturbation theory and the Dirac relativistic framework. \\
Results for the 
angular correlation have been showed for the transitions $2s_{1/2}\,(F=1,0)\to1s_{1/2}\,(F=1,0)$ in Hydrogen atom as well as for
$2s_{1/2}\,(F=4,3)\to1s_{1/2}\,(F=4,3)$ in hydrogenlike ${}^{235}_{92}$U ion. It has been possible to identify 
two types of emission pattern: 1- the two-photon decays which connect
ionic states with the same spin (non-spin-flip transitions) are found to be characterized by an angular correlation 
$W(\theta)\sim1+\cos^2\theta$, with small deviations which are increasing with the atomic number Z; 2- the two-photon decays which connect
ionic states with different spin (spin-flip transitions) are found to have much smaller decay rate and to be characterized by an angular correlation
$W(\theta)\sim1-1/3\cos^2\theta$, with deviations which are already important for low-Z systems.
In both cases, the (asymmetric) deviations come
from high-multipole contributions. \\
Results for the polarization correlation have been showed for the transitions $2s_{1/2}\,(F=0)\to 1s_{1/2}\,(F=1,0)$ in
Hydrogen as well as for $2s_{1/2}\,(F=3)\to 1s_{1/2}\,(F=4,3)$ in hydrogenlike ${}^{235}_{92}$U ion.
It has been found that the photons coming from the studied F = 0 $\to$ F = 0 transition in Hydrogen atom are exclusively characterized
by parallel linear polarization.
By using such results,
it has been possible to outline a way to experimentally measure the 
angular and polarization correlations in the transition $2s_{1/2}\,(F=0)\to 1s_{1/2}\,(F=1)$, in Hydrogen atom.
\begin{acknowledgments}
The author acknowledges Prof. R.B. Thayyullathil for his suggestions.
\end{acknowledgments}


\newpage

\end{document}